\documentclass[twocolumn,aps,pra]{revtex4}
\usepackage{epsfig}
\usepackage{amsmath}
\begin{document}

\title{Quantitative study of desctructive quantum interference effect on the lin$\|$lin CPT.}
\author{E. Breschi$^1$\thanks{E--mail:
evelina.breschi@unine.ch}, G.Kazakov$^{2}$, R. Lammegger$^3$,
G.Mileti$^{1}$, B.Matisov$^{2}$, L. Windholz$^3$}
\affiliation{{\setlength{\baselineskip}{18pt}
$^1${Laboratoire Temps-Fr$\acute{e}$quence - University of Neuch$\hat{a}$tel,
2000 Neuch\^{a}tel, Switzerland}\\
$^{2}${St. Petersburg State Polytechnic University,
St. Petersburg, 195251, Russia}\\
$^{3}${Institute of Experimental Physics TU-Graz, 8010 Graz, Austria}\\
}}

\begin{abstract}
We investigate experimentally and theoretically the Coherent Population Trapping
(CPT) effect occurring in $^{87}$Rb D$_1$ line due to the interaction with linearly
polarized laser light (lin$\|$lin CPT). In this configuration the coherence is strongly influenced by the structure of the excited
state; consequently the quantum interference between dark states is an essential feature of
this interaction scheme. We study the lin$\|$lin CPT resonance as a function of the laser optical
detuning. The comparison between experimental 
 theoretical results allows us to
quantify the contribution from different dark states to the total signal.
Based on these results we investigate the signal depending on both the pressure broadening
of the optical transition and the laser linewidth, and we find in which conditions the
laser linewidth does not degrade the lin$\|$lin CPT resonance.\\
\vspace*{10pt}
{PACS numbers: }
\end{abstract}
\maketitle
\section{Introduction}
\label{intro} The first investigations of the Coherent Population
Trapping (CPT) effect were performed theoretically and
experimentally in the seventies \cite{Alzetta76}. In accordance with
these early works, the CPT effect is due to a coherence between
ground states caused by the interaction with a quasi resonant,
two-frequency and coherent light field. CPT and related effects
(such as Electromagnetically Induced Transparency \cite{Harris97}
and Absorption \cite{Lezama99}) have an impact on both, the atomic
system state and the light beam propagation through the media.

CPT resonances with a narrow linewidth can be recorded in the simplest case in
quantum systems with two long-lived ground states and one excited
state coupled by a dichromatic coherent light field.
In this configuration the quantum system is prepared in a non absorbing state (the dark state).
The energy levels in such a configuration are arranged similar to the Greek letter
$\Lambda$ (so called $\Lambda$-System) and can be found, for
instance, in alkali atoms. In the experiment CPT resonances are
recorded by varying the frequency difference of two spe\-ctral
electromagnetic field components around the value of the
frequency splitting of the ground states. The properties of the CPT
resonances can be exploited in a wide range of applications such as
atomic magnetometry \cite{Nagel98}, atomic frequency standards
\cite{Vanier05}, pulse delaying and compression for optical memory
\cite{Hau99}.

The problem in the use of CPT effect in those applications is its low strength:
often the dark state induces only a small variation in the light trasmitted throught the atomic sample.
To overcome this limitation recently se\-ve\-ral novel
light-atom interaction schemes were proposed
\cite{Rosenbluh06,Kargapoltsev04}. In particular in work
\cite{Taichenachev05} the authors show that a significant
enhancement of the CPT effect can be obtained in the so-called
lin$\|$lin CPT configuration. Here two co-propagating linearly
polarized laser waves with parallel linear polarization vectors are
resonant with the transitions to $5\,^2$P$_{1/2}$ F$_e=1$ of the
$^{87}$Rb isotope. A detailed analysis of the two-photon $\Lambda$ transition
process shows that in a lin$\|$lin CPT configuration the CPT resonance depends critically on the excited state
hyperfine structure (HFS) \cite{Wynands98}, in a way different to that of the well
known CPT resonance obtained by a circularly polarized laser field \cite{Knappe03}.
In the case of circularly and linearly polarized laser field the dark state arise from a vectorial ($\Delta m_F$=0)
and quadrupolar coupling ($\Delta m_F$=2), respectively.

A theoretical study of four-level system describing the
CPT resonance depending on the structure of the
ground and excited atomic levels has been published in
\cite{Lukin99}. The interest of these systems relies to the
possibility of controlling its optical properties via the quantum
interference arising between two dark states, for instance, by means
of the phases of the laser fields \cite{Kosachev92,Korsunsky99} or
the local phase of the dark states \cite{Affolderbach02}. In the
present work we study the behavior of the lin$\|$lin CPT
resonance occurring within the manifold of the hyperfine
transitions of the $^{87}$Rb D$_1$ line. In our case
the CPT resonance depends on the structure of the excited hyperfine
states and it is described in detail in section \ref{sec:model} with
the model developed for the data interpretation. We analyze the lin$\|$lin
 CPT resonances as a function of the laser detuning and the
homogeneous broadening (in section \ref{sec:result1}), which depends
on the pressure broadening (due to the buffer gas in the cell) and
the laser linewidth. Note that the evaluation of the laser linewidth
influence on the dark states is crucial for applications and has not been
investigated much so far. A theoretical study of the laser linewidth
effects on CPT resonance for achieving selective excitation of atoms (with
application in laser cooling and quantum computing) is presented in
reference \cite{McDonnell04}; while in \cite{Yavuz07} the author has
demonstrated with his model a scheme that allows large time delay
for large bandwidth optical pulse.

We perform our studies with two different light
sources: a pair of Phase-Locked (PL) Extended Cavity Diode Lasers (ECDLs) and a current
modulated Vertical Cavity Surface Emitting Diode Laser (VCSEL).
The PL-ECDLs have, in good approximation, pure dichromatic fields
with a narrow li\-ne\-width. On the contrary, the modulated VCSEL
has a multi-frequency spectrum with broader linewidth. A detailed
discussion of the laser sources and the experimental setup is
presented in section \ref{sec:setup}.
\subsection{Application of the lin$||$lin CPT resonance in compact atomic clocks}
\label{sec:clock}
\begin{figure}
  \begin{center}
    \resizebox{0.45\textwidth}{!}
    { \includegraphics{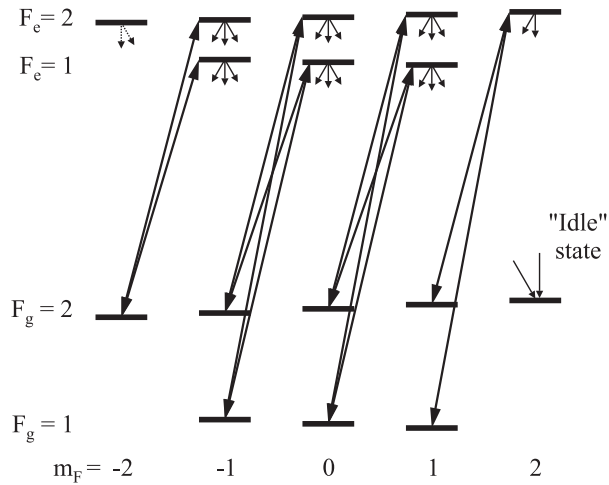}}
  \end{center}
  \caption{Schematic diagram of the light induced transitions giving the
  $\sigma^+$-$\sigma^+$ CPT resonance by excitation of the hf
  transitions belonging to the D$1$ line of the $^{87}Rb$ isotope.
  The $\Lambda$-system involving the ground states $|F_g=1, m_F=0\rangle$
  and $|F_g=2, m_F=0\rangle$ gives rise to the (often called) 0-0
  CPT resonance wich is used as reference in atomic clocks.
  The atomic population is partially collected in the state
  $\left|F_g=2, m_F=2\right\rangle$ via an optical
  pumping process. For this reason the state  $\left|F_g=2, m_F=2\right\rangle$
  is called \emph{idle state}. The small arrows represent the spontaneous decay of excited
  states.}
  \label{fig:sigma-sigma-CPT}
\end{figure}
In the last years many efforts has been steered into the realization of compact CPT-based
atomic clocks \cite{Knappe08}. The CPT
excitation scheme usually used is based on the interaction of Cs or
Rb atoms with a circularly polarized laser field, i.e. either
$\sigma^+$ or $\sigma^-$ transitions are induced by the two
frequency components of the driving laser field (in the following called
$\sigma$-$\sigma$ CPT). A typical excitation scheme in case of the
manifold of hyperfine transitions within the $^{87}$Rb D$_1$ (the same
case we will discuss for the lin$||$lin configuration) is illustrated in figure
\ref{fig:sigma-sigma-CPT}. Here, a weak magnetic field of few tens of
$\mu$T oriented in laser propagation direction, is applied to lift
the degeneracy of the Zeeman sublevels. In this configuration the resonance used
 as reference in frequency standard applications is the coherent
superposition of the sublevels $|F_g=1, m_F=0\rangle$ and $|F_g=2,
m_F=0\rangle$ being in first order not sensitive to magnetic fields.

The problem of this interaction scheme relies on the fact that
the state $|F_g=2, m_F=2\rangle$ is not involved in the
excitation process. As a consequence, the optical pumping effect caused by the
isotropic spontaneous decay of the excited states accumulates the
population into this \emph{idle state} which, thus, represents a loss
mechanism for the clock reference resonance. This leads (especially at higher laser intensities and low buffer gas pressure) to a
reduced signal/noise ratio because most of state population is
concentrated in those Zeeman sublevels with highest (lowest) $m_F$
quantum numbers.

On the contrary, in case of a lin$||$lin CPT configuration
(discussed in section \ref{sec:model}) no idle states are present
and the state population is concentrated symmetrically around the
Zeeman sublevel with quantum number $m_F=0$. This can be understood
qualitatively because there is no transfer of net angular momentum
from the li\-near po\-la\-ri\-zed light to the atoms in the
excitation process. Thus a better signal-to-noise ratio is expected
especially in the case of higher laser intensities and low buffer gas
pressures. This characteristic makes the lin$||$lin CPT resonance a
valid candidate for high performance compact atomic clock (a
detailed study is presented in \cite{Breschi}).

\section{Model and interaction scheme}
\label{sec:model}
%
%
A model based on the density matrix approach has been used for the
data interpretation and has been described in detail in
reference \cite{KazTechPhys06}. Here we report on the basic
approach and we give an intuitive picture of the results essential
for our discussion.

Let us to consider a $^{87}$Rb atom excited by a two-frequency laser
field resonant to hf-transitions $5\,^2S_{1/2}$ $F_g=1,2$
$\leftrightarrow$ $5\,^2P_{1/2}$ $F_e=1$ of the $D_1$ line. For such
a system the density matrix evolution is given by
\begin{equation}
\dot{\rho}_{ij}= -\frac{i}{\hbar}
\sum_k[H_{ik}\rho_{kj}-\rho_{ik}H_{kj}]+\sum_{k,l}\Gamma_{ij,kl}\rho_{kl}.
\label{eq:DensityMatrix}
\end{equation}
The Hamiltonian, $\widehat{H}$ has two terms: the atomic
un-perturbed Hamiltonian, $\widehat{H}_0$, and the interaction operator,
$\widehat{V}$, i.e. $\widehat{H}=\widehat{H}_0+\hbar\widehat{V}$.
$\Gamma_{ij, kl}$ is the relaxation matrix element describing the
relaxation processes.

The two frequency laser field can be written as follow:
\begin{equation}
\vec{E}(z,t) = \sum_{j=1,2}
[\frac{\vec{E}_j}{2}\exp[i(\omega_jt+\varphi_j(t)-k_j z)] +c.c.,
\label{eq2}
\end{equation}
whereas the laser line width is modeled by the phase fluctuations
$\varphi_j(t)$.

The absorbed laser power in a optically
thin gas cell (weak absorption) can be calculated by:
\begin{equation}
\Delta P = \rho_{exc} \hbar \omega_{opt} \gamma N_{a}, \label{eq2a}
\end{equation}
where $\rho_{exc}$ is the total excited state population,
$\omega_{opt}$ is the optical transition frequency, $\gamma$ is the
excited state relaxation rate, and $N_{a}$ is the number of active
atoms taking part in the excitation process. In a optically thin gas
cell, due to a weak absorption, $\rho_{exc}$ is not a function of
the optical path because the matrix elements of the interaction
Hamiltonian $\hbar V_{ij}$ are assumed to be constant along the
optical path. At high temperatures  the gas cell becomes optically
thick. The consequence is that the density matrix elements (and with
them the dark states) are functions of the light intensity along the
optical path. Modeling of such a coupled behavior is rather
complicated because, strictly speaking, the effects of propagation
determined by the Maxwell equations are strongly coupled with the
density matrix equations of the quantum system. Our experiments are
performed with an optically thick $^{87}$Rb vapor cell.
Therefore in the model the optical path is divided in a set of
subsequent parts. Assuming constant electrical fields ($E_i=const.$)
in each thin layer the density matrix is calculated. Finally the
Maxwell equations for the slowly varying E-field amplitude and phase
\cite{Kosachev95}, are used to calculate the variations of the
E-fields within the layer. Thereafter the density matrix of the next
layer is determined by using the new E-field derived from the
previous layer. This procedure is repeated for all subsequent parts.
In the numerical calculations we don't take into account
modifications of the transverse intensity distribution.
Nevertheless, this simple phenomenological approach allows us to
reproduce the experimental results (see Section \ref{sec:result1}).
In the following the ground and excited state Zeeman sublevels are
called $g$, $g'$ and $e$, respectively. Note that $g$ and $g'$
belong to different hyperfine ground states and a dark state can be
created when the two transitions, $\left| g \right\rangle
\leftrightarrow \left|e\right\rangle$ and $\left| g'\right\rangle
\leftrightarrow \left| e \right\rangle$, are excited simultaneously.
If the laser field intensity is much lower than the sa\-tu\-ra\-tion
limit for each optical transition and the cell is optically thin,
the total excited state population $\rho_{exc}$ is proportional to
the light power absorbed (c.f. equation \ref{eq2a}). Neglecting fast
oscillating terms $\sim\mathcal{O}(\omega_{i}+\omega_j)$ and
$\sim\mathcal{O}(2\omega_j)$ etc. (rotating wave approximation) the
excited state density matrix element is connected via
\begin{equation}
\rho_{exc}=\!\sum_{e,g,g'}\left[ \frac{V^{0}_{eg}V_{g'e}}{\gamma'
\gamma} (G_{ge}+G_{g'e}+i(F_{ge}-F_{g'e})) \rho_{gg'}\right]
\label{eq:RhoExc}
\end{equation}
to the ground state density matrix elements which can be derived
from the following set of equations:
\begin{equation}
\label{eq:Rho}
\begin{split}
\dot{\rho}_{gg'}=&-i
\rho_{gg'}(\omega_{gg'}-\omega_{g}+\omega_{g'})+
\Gamma\,(\frac{1}{8}\,\delta_{gg'}-\rho_{gg'})-\\
\sum_{g'',e} &\left[\frac{V_{ge}V_{eg''}}{\gamma'}\,(G_{g'}+iF_{g'e})\right]\rho_{g''g'}-\\
\sum_{g'',e}&\left[\frac{V_{g''e}V_{eg'}}{\gamma'}\,(G_{ge}-iF_{ge})\right]\rho_{gg''}+
\frac{\delta_{gg'}}{8}\boldsymbol{\cdot}\\
\sum_{e,g'',g'''}&\left[\frac{V_{g''e} V_{eg'''}}{\gamma'}
\,(G_{g''e}+G_{g''e}+i(F_{g'''e}-F_{g''e}))\right]
\end{split}
\end{equation}
\begin{equation}
\sum_{g}\rho_{gg}=1. \label{eq:Norm}
\end{equation}
Here $\hbar V_{ij}$ are the matrix elements of the interaction
Hamiltonian (Rabi frequencies) in the frame rotating with the
corresponding laser frequency component; $\gamma'$ is the decay rate
of the optical coherence $\rho_{eg}$; $\delta_{ gg'}$ is the
Kronecker delta. $\omega_g$ is the frequency of laser component
interacting with the level $g$; and $\omega_{gg'}$ is the frequency
spacing between the levels $|g\rangle$ and $|g'\rangle$. The ground
state relaxation rate denoted by $\Gamma$ depends on the
temperature of the cell, type and pressure of buffer gas, geometry
of the cell, the laser power and the geometry of laser beam\cite{Violino}. 
In our simulation $\Gamma$ is the free fit parameter
in our simulation which is estimated from the measurements.

In \cite{Mazets92}, the authors demonstrate that the optical
coherence decay rate $\gamma'$ is determined by the laser linewidth
$\Gamma_{L}$, by the spontaneous relaxation rate $\gamma_{sp}$ which
is for Rubidium $\simeq 2 \pi\cdot$ $5.6$ MHz, and by the pressure
broadening $\gamma_{c}$ which depends on the type of buffer gas
and on the buffer gas pressure in the cell:
\begin{equation}
\gamma'=\frac{\gamma_{sp}+\gamma_{c}+\Gamma_{L}}{2}.
\label{eq:gammaFirst}
\end{equation}

\begin{figure}
  \begin{center}
    \resizebox{0.4\textwidth}{!}
    { \includegraphics{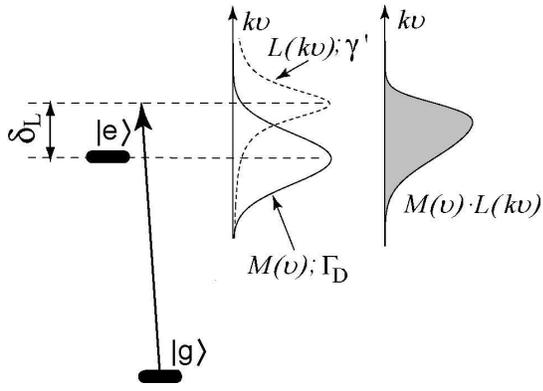}}
\end{center}
  \caption{Schematic view of the homogeneous broadening profile
    (Lorentzian curve $L(kv)$ of linewidth $\gamma'$ in dashed line), of a transition
($\left|g\right\rangle \leftrightarrow \left|e\right\rangle$).
$\delta^{eg}_L$ is the laser detuning. The
    Gaussian curve (of linewidth $\Gamma_D$) in solid line, represents
    the Doppler profile $M(v)$ of the transition.
    In equation (\ref{eq:GF}) G$_{eg}$ is proportional to the grey area under the
product of the two curves ($L(kv) \cdot M(v)$).}
  \label{fig:hombrod}
\end{figure}

To understand the physical meaning of $\gamma'$ let us consider a
generic single transition $\left|g\right\rangle \leftrightarrow
\left|e\right\rangle$. The excitation of this transition by a
laser light field of linewidth $\Gamma_L$, results in an
homogeneous broadened profile that can be described with a
Lorentzian curve of linewidth $\gamma'$ (figure
\ref{fig:hombrod}).

The coefficients $G_{ge}$ and $F_{ge}$ in equation \ref{eq:RhoExc}
and \ref{eq:Rho} are real and equal to:
\begin{equation}
\begin{split}
G_{ge}&= \int_{-\infty}^{+\infty} \frac{(\gamma')^2
  M(v)}{(\gamma')^2 + (\delta_L^{eg}-kv)^2} dv\\
F_{ge}&= \int_{-\infty}^{+\infty} \frac{\gamma'(\delta_L^{eg}-kv)
  M(v)}{(\gamma')^2 + (\delta_L^{eg}-kv)^2} dv
\label{eq:GF}
\end{split}
\end{equation}
where $M(v)$ is the atomic velocity distribution and
$\delta_{L}^{eg}$ is the laser detuning.
The coefficient G$_{ge}$  is proportional to the strength of the
single optical transition $\left|g\right\rangle \leftrightarrow
\left|e\right\rangle$; while the coefficient F$_{ge}$ allows to
calculate the shift of the resonance frequency. For our intent we
focus our attention to the coefficient G$_{ge}$ which is
schematically represented in figure \ref{fig:hombrod}. $G_{ge}$ is
proportional to the grey area delimited by the product of two profiles: the
Doppler profile, represented by a Gaussian curve $M(v)$ of linewidth
$\Gamma_D$, and the homogeneous profile, represented by a Lorentzian
curve $L(kv)$ of linewidth $\gamma'$. If different $\left|g\right\rangle\leftrightarrow\left|e\right\rangle$ transitions are
excited, G$_{ge}$ characterizes the contribution of
each transition to the total excitation process.

Now we extend the previous consideration made for a single optical
transition to a dark state. The necessary condition for creating the
dark state is that the transitions from both ground states $\left| g
\right\rangle$ and $\left|g'\right\rangle$ towards the same excited
level $\left|e\right\rangle$ are simultaneously excited by two
coherent electromagnetic field components, i.e. the relative
frequency jitter of the electromagnetic field components is
negligible. This condition is fulfilled either in the VCSEL and in the
PL-ECDLs (see Section \ref{sec:setup}). The straightforward
consequence is that the one-photon detuning of both light field
components must coincide ($\delta_L^{eg}=\delta_L^{eg'}$). Under
this condition, from equation \ref{eq:GF} the relation G$_{ge}$=
G$_{g'e}=$G$_e$ can be derived. In synthesis the G-coefficient for a
general dark state depends only on the excited states involved (and
not on the ground states). When different dark states occur simultaneously,
the contribution to the total signal of each single dark state can be characterized by calculating G$_e$.
Our model takes into account the Zeeman and hyperfine structure of
the $^{87}$Rb atoms. In the following we define the interaction
scheme, i.e. which group of atomic sublevels are involved in the dark state
preparation. Finally we present the results obtained by applying the model to
an isolated reduced system, in order to show how the G$_e$
coefficients describe the CPT resonance strenght.
%
%
\begin{figure}
  \begin{center}
    \resizebox{0.45\textwidth}{!}
    { \includegraphics{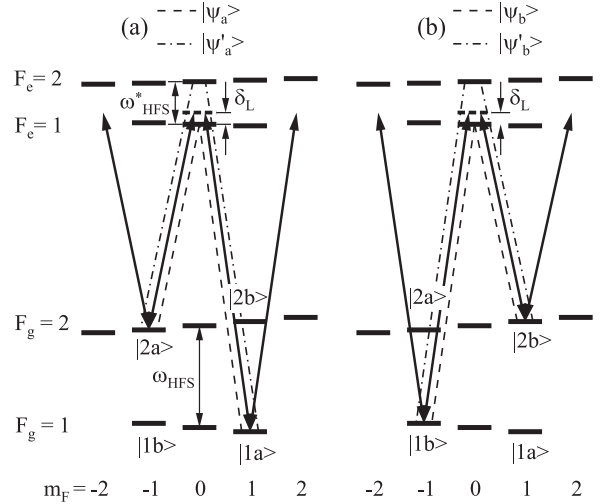}}
\end{center}
  \caption{Schematic diagram of the light induced transitions
  giving the lin$\|$lin CPT resonance at $\omega\approx\omega_{HFS}$. The levels involved are:
    $\left|1 a\right\rangle=\left|F_g=1,m_F=+1\right\rangle$
    and $\left|2 a \right\rangle=|F_g=2,m_F=-1\rangle$ part
    (a) of the figure; $\left|1 b\right\rangle=\left|F_g=1,m_F=-1\right\rangle$
    and $\left|2 b\right\rangle=|F_g=2,m_F=+1\rangle$ part (b) of the figure.
    In presence of a low magnetic field along the laser
    propagation direction (=quantization axis) the dark states
    in (a) and (b) are degenerate. The solid arrows are representing the
    components of the exciting laser field and $\delta_L$ is the laser detuning. }
  \label{fig:linlin}
\end{figure}

Figure \ref{fig:linlin} shows the relevant transitions induced by
the coherent $\sigma^+$ and $\sigma^-$ components of the linearly
polarized light fields expressed in spherical tensor basis. In
figure \ref{fig:linlin} we do not report the $\sigma^{+}-\sigma^{+}$
and $\sigma^{-}-\sigma^{-}$ groups of transitions. In general, the
existence of a CPT resonance is directly connected via the phase
relation between the Rabi frequencies (phase relation of laser
fields and atomic wave functions) of each transition of the
$\Lambda-$scheme \cite{Kosachev92}. In the specific case of
lin$\|$lin excitation the dark states arising from the
$\sigma^{+}-\sigma^{+}$ and $\sigma^{-}-\sigma^{-}$ are orthogonal,
thus they interfere destructively and do not contribute to the CPT
resonance at $\omega\approx \omega_{HFS}$ \cite{Rosenbluh06}.

We study the $\sigma^{+}-\sigma^{-}$ transitions between ground
state Zeeman sublevels with $m_F=\pm1$.  The feature of the CPT
resonance is related with the presence of two excited levels
(F$_e$=1,2), both allowed for dipole transitions. In synthesis, four
dark states contribute to the lin$\|$lin CPT resonance at
$\omega\approx \omega_{HFS}$: $|\Psi_a \rangle$ and $|\Psi_b\rangle$
are the coherent superpositions of the Zeeman sublevels $|1a\rangle
\leftrightarrow |2a\rangle$ and $|1b\rangle \leftrightarrow
|2b\rangle$ through the excited state F$_e=1$, while $|\Psi'_{a}
\rangle$  and $|\Psi'_{b} \rangle$ are the coherent superpositions
of the same Zeeman sublevels via the excited state F$_e=2$. When a
magnetic field is applied along the quantization axis $|\Psi_{a}
\rangle$-$|\Psi_{b}\rangle$ and $|\Psi'_a \rangle$-$|\Psi'_b\rangle$
split with a factor of $\pm 28$ Hz$/\mu$T determined by the nuclear
g-factor. Remark that the transitions towards the outermost Zeeman
sublevels $\left|F_e=2, m_F=\pm2\right\rangle$, play an important
role in the formation of the dark state $|\Psi'_a \rangle$ and
$|\Psi'_b \rangle$, as we are going to show in next paragraph.

\subsection{Simplified atomic system: analytical solution}
\begin{figure}
  \begin{center}
    \resizebox{0.45\textwidth}{!}
    { \includegraphics{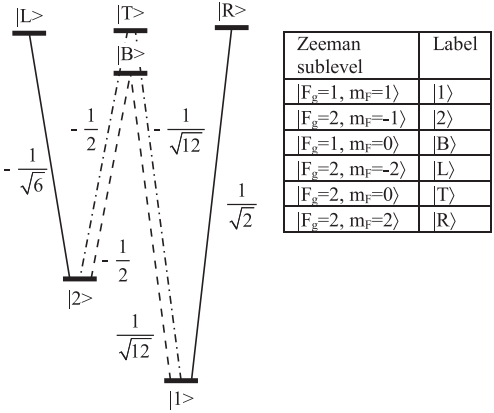}}
\end{center}
  \caption{Isolated 6-level excitation system. The coefficients reported
  near the transition are proportional to the Rabi frequencies $V_{eg}$, i.e.
  they are the dipole matrix element for the transitions expressed as multiples
  of the coefficient
  $V^0=\sqrt{\frac{3\gamma_{sp} c^3 \hbar}{2\omega^3}}$.}
  \label{fig:simply}
\end{figure}

%
%
To point out how the G$_e$ coefficients can describe the
lin$\|$lin CPT resonance we apply the approach presented at the
beginning of this section to figure \ref{fig:linlin} (a), under the
hypothesis that the 6 levels participating to the interaction are
isolated. Similar considerations can be applied to figure
\ref{fig:linlin} (b) and then to the lin$\|$lin CPT resonance.
Figure \ref{fig:simply} is the diagram of the isolated 6-level
system. The Zeeman sublevels are named with a short label for
handling with a compact notation. The coefficients reported near
each transition are the dipole matrix elements expressed as
multiples of $V^0=\sqrt{\frac{3\gamma_{sp} c^3 \hbar}{2\omega^3}}$.
These coefficients are proportional to the Rabi frequencies,
$V_{eg} $\cite{VarMosKhers,LL}. For instance, in our notation the
Rabi frequency for the transition $\left|1\right\rangle
\leftrightarrow \left|R\right\rangle$ is named $V_{1R}$ and is equal
to $(\frac{1}{\sqrt{2}}\cdot\frac{V^0\cdot E_1^+}{2\hbar})$, where
$E_1^+$ is a circularly po\-la\-ri\-zed component of $\vec{E}_1$.
There are only two ground state not degenerate, $|1\rangle$ and $|2\rangle$,
in our simplified scheme, therefore the normalization condition
(\ref{eq:Norm}) in the regime of low laser field intensities
($I_{las}\ll I_{sat}$) gives $\rho_{11}+\rho_{22}\simeq1$. 
There are two excited states, thus $e=1,2$ such as for $e=1$ the excited state has one sub-level, $|B\rangle$, and for $e=2$ the excited state has three sub-levels 
$|L\rangle$, $|T\rangle$ and $|R\rangle$.
We introduce for simplicity the new variables: $f$ (population
difference), $R$ and $J$ (real and imaginary part of $\rho_{12}$):
\begin{equation}
\rho_{11}=(1+f)/2 ,\quad \rho_{22}=(1-f)/2,\quad \rho_{12}=R+i \cdot J.
\label{eq:fRJvar}
\end{equation}

The total excited state population $\rho_{exc}$ can be obtained by
applying equation (\ref{eq:RhoExc}):
\begin{equation}
\rho_{exc}=\frac{1}{\gamma} [W+(W_1-W_2) \cdot f+4W_{12} \cdot R].
\label{eq:rhoExcSimple}
\end{equation}
From equations (\ref{eq:Rho}) and (\ref{eq:fRJvar}), we obtain the
set of equations for variables $f$, $R$ and $J$:
\begin{eqnarray}
\label{eq:fRJeqs}
\dot{f}&=&-(\Gamma+W) \cdot f-4 D_{12} \cdot J+W_2-W_1, \nonumber \\
\dot{R}&=&-(\Gamma+W) \cdot R+(\Omega-\Delta)\cdot J- W_{12},  \\
\dot{J}&=&D_{12} \cdot f-(\Gamma+W) \cdot J-(\Omega-\Delta) \cdot R. \nonumber
\end{eqnarray}
Here $\Omega=(\omega_{2}-\omega_{1}-\omega_{21})$ is the Raman
detuning and $\Gamma$ is ground state relaxation rate. $W$ is the
so-called "optical-pumping rate" and the quantity $\Delta$ is a
measure for the light shift of the micro-wave transition (i.e.
the shift of $\omega_{21}$, i.e. the frequency difference between
$|1>$ and $|2>$ in figure \ref{fig:simply}). Following our notation we obtained:
\begin{eqnarray}
&W_1&=\frac{|V_{1B}|^2}{\gamma'}G_1+\frac{|V_{1T}|^2+|V_{1R}|^2}{\gamma'}G_2,
\label{eq:W1} \\
&W_2&=\frac{|V_{2B}|^2}{\gamma'}G_1+\frac{|V_{2T}|^2+|V_{2L}|^2}{\gamma'}G_2, \label{eq:W2}\\
&W&=W_1+W_2, \label{eq:W}\\
&W_{12}&=\frac{V_{1B} V_{B2}}{\gamma'}G_1+\frac{V_{1T} V_{T2}}{\gamma'}G_2, \label{eq:W12}\\
&D_{12}&=\frac{V_{1B} V_{B2}}{\gamma'}F_1+\frac{V_{1T} V_{T2}}{\gamma'}F_2, \label{eq:D12}\\
&\Delta&=\frac{|V_{2B}|^2-|V_{1B}|^2}{\gamma'}F_1+  \nonumber \\
&&
\frac{|V_{2T}|^2+|V_{2L}|^2-|V_{1T}|^2-|V_{1R}|^2}{\gamma'}F_2.
\label{eq:Delta}
\end{eqnarray}
Where $G_1$ and $G_2$ are the $G$-coefficients for $e=1,2$ determined by applying eq. \ref{eq:GF} to the simplified 6-level scheme.
Substituting the stationary solution of the set of equations
(\ref{eq:fRJeqs}) into (\ref{eq:rhoExcSimple}), we obtain the
following analytical expression for the total excited state
population:
\begin{equation}
\begin{split}
\rho_{exc}=&\frac{1}{\gamma}\biggl[W-\frac{(W_1-W_2)^2+4W_{12}^2}{\Gamma
+
W}+\\
&\frac{4(D_{12}(W_1-W_2)-W_{12}(\Omega-\Delta))^2}{(\Gamma+W)
((\Gamma+W)^2+4D_{12}^2+(\Omega-\Delta)^2)}\biggr]
\label{eq:rhoExcSolution}
\end{split}
\end{equation}
The first two terms in (\ref{eq:rhoExcSolution}) do not depend on
the two-photon detuning $\Omega$; the third term gives the change in
the absorption due to the CPT effect. Finally, we substitute the
Rabi frequencies ($V_{eg}$) into (\ref{eq:W1}), (\ref{eq:W2}) and
(\ref{eq:W12}) and we get:
\begin{eqnarray}
&&W_1=\frac{|V^0|^2}{\gamma'}\left[\frac{1}{12}G_1+\frac{7}{12}G_2\right],\label{eq:W1res} \\
&&W_2=\frac{|V^0|^2}{\gamma'}\left[\frac{3}{12}G_1+\frac{5}{12}G_2\right], \label{eq:W2res}\\
&&W_{12}=\frac{|V^0|^2}{\gamma'}\frac{G_2-G_1}{4\sqrt{3}},
\label{eq:W12res}
\end{eqnarray}

When $G_1=G_2$, the two terms ($W_1-W_2$) and $W_{12}$ are
vanishing, therefore the third term in \ref{eq:rhoExcSolution} is
vanishing, and the CPT resonance goes to zero. This is an evidence
of destructive quantum interference between the different dark states
prepared through the two excited state hyperfine
sublevels.

In this section we summarize a method for calculating the CPT resonance
based on the fact that the light power absorbed by the atoms is proportional to the detected signal
and can be calculated by using equation \ref{eq2a}, \ref{eq:RhoExc} and \ref{eq:Rho}.
On the basis of a simplified 6-level system, we showed analytically that the
excited state hyperfine structure of $^{87}$Rb plays an important role in the CPT
excitation process due to a destructive quantum interference effect. It is found that the
characteristic of this interference can be well described by the
ratio $G_1/G_2$ determined by the laser detuning $\delta_L$ and the
optical coherence decay rate $\gamma'$. A ratio close to unity
expresses similar (excitation) strengths in both
dark states which leads to a high degree of
interference and to a cancelation of the CPT resonance.
\section{Experimental setup}
\label{sec:setup}
\begin{figure}
  \begin{center}
    \resizebox{0.45\textwidth}{!}
    {\includegraphics{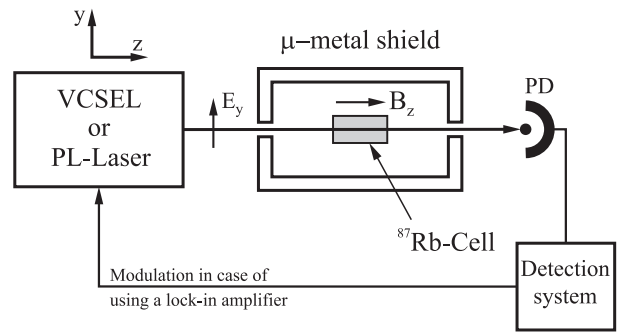}}
\end{center}
  \caption{Schematic block diagram of the experimental setup.
  Two laser light sources have been alternatively used during the experiments:
  a modulated VCSEL and two PL-ECDLs.
  The detection system block consists of a current amplifier,
  a digital oscilloscope read out by a PC and a lock-in amplifier (optional). 
  E$_y$ indicates the direction of the electrical field vector of the linearly polarized laser radiation.
  In the experiments a magnetic field B$_z$= 3 $\mu$T is applied in laser propagation direction.}
\label{fig:setup}
\end{figure}

%
%
A sketch of the experimental setup is shown in figure
\ref{fig:setup}. The core is a glass cell with a volume of a few
cm$^3$ containing the $^{87}$Rb isotope and N$_2$ as buffer gas. In
particular two similar cells are used containing $0.5$ and $1.5$ kPa
of buffer gas, respectively. The cell temperature during is stabilized to ($68 \pm 1)^\circ$C, corresponding to a
Doppler broadening of $\Gamma_D= 2 \pi\cdot540$ MHz of the optical
transitions. Since the light fields are both co-propagating the
Doppler broadening of the CPT resonance is due to the $6.8$ GHz
difference frequency. Note, that - in the case of a buffer gas cell
- the micro-wave Doppler effect is strongly reduced by the Dicke
narrowing \cite{Dic53} because the atoms are confined within a
volume much smaller than the $4.4$ cm wavelength corresponding to
the $6.8$ GHz frequency. At a temperature of $68^{\circ}$C the
$^{87}$Rb cell becomes optically thick. Therefore, when a single
mode laser (VCSEL or ECDL) is in resonance, the cell transmittance
(i.e. the ratio between the laser intensity after and before the
cell ($I/I_0$)) is always $\leq 0.3$ corresponding to an optical
thickness $\geq1$.

The laser radiation transmitted through the cell is collected onto a
photodetector (PD) whose signal is amplified in a
current amplifier afterwards. Finally, the variation (increase) of
the optical transmission is recorded by an digital oscilloscope
directly connected to the current amplifier. Additionally, the CPT
resonance can be monitored by using a lock-in amplifier.

In order to avoid the influence of spurious magnetic fields and
magnetic gradients, the cell was inserted in a CO-NETIC alloy
magnetic shield. Inside the shiel\-ding, a solenoid provided the
longitudinal magnetic field, B$_z=(3.0\pm0.2)$ $\mu$T in order to
lift the degeneracy of the Zeeman sublevels. The magnetic field has
been re\-gu\-larly monitored during the experiments by using the g$_J$
dependence of the outermost CPT resonance \cite{Taichenachev05}, i.e.
the dark states created by the coherence of $\left| F_e=1, m_F=0
\right\rangle$ and $\left|F_e=2, m_F=\pm2\right\rangle$.
The functional block \emph{VCSEL or PL-Laser} indicates that the
experiments are performed either with a
current modulated \emph{VCSEL} or with \emph{PL-ECDLs}. Both laser system have a Gaussian
beam profile and a beam waist of $2$ mm ($1/e^2$). However, it is
important to outline that the relevant difference between both laser
systems is their spectral linewidth $\Gamma_V>100\cdot\Gamma_{PL}$.
The characteristic of the two laser systems are shortly described in
the following.

\emph{Modulated VCSEL} (see ref \cite{Affolderbach00}): The
injection current of a single mode VCSEL emitting at $795$ nm, is
directly modulated with $3.417$ GHz (i.e. half of ground state
hyperfine separation in $^{87}$Rb ) with 10 dBm of RF power. The
VCSEL has a broad spectrum, the measured linewidth is
$\Gamma_V\approx2\pi\cdot100$ MHz. The modulation performance has
been evaluated through the one-photon $^{87}$Rb absorption spectrum
by changing the frequency and the amplitude of the current
modulation. We did not notice any influence of amplitude modulation,
and the frequency modulation index was evaluated to be about $1.8$.
Under these conditions about $68\%$ of the total laser power is
equally distributed in the first-order side-bands which are used for
dark state excitation. The remaining $32\%$ of the total power is
distributed among mainly the carrier frequency and the higher-order
side-bands. These off-resonant parts of the VCSEL-spectrum can cause
one-photon excitation processes because of the Doppler broadening of
the hyperfine transitions. Additionally, the not absorbed
off-resonant light is increasing the dc and shot-noise levels of the
photo-detector.
\begin{figure}
    \begin{center}
        \resizebox{0.45\textwidth}{!}
        {\includegraphics{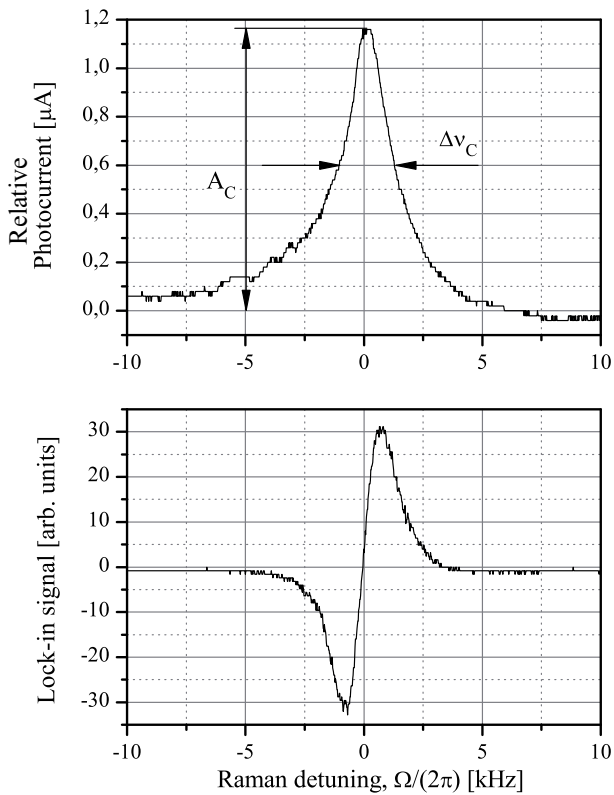}}
    \end{center}
\caption{An example of a typical CPT resonance prepared with
PL-ECDLs resonant with the group of transitions towards $F_e=1$
(laser intensity equal to $3.8$ $mW cm^{-2}$). Both the oscilloscope
(top) and lock-in (bottom) signals are represented. The amplitude of
the CPT resonance and its line widths are $A_C=1$ $\mu$A and $\Delta
\nu_C=3$ kHz, respectively. The variation of the normalized optical
transmission due to the CPT is about $4\%$. Note that in the same
condition with modulated VCSEL we record about $3\%$ of normalized
optical transmission variation. The difference is due to the
background light level increased by the presence of the
off-resonance frequencies.} \label{fig:lockin-dc}
\end{figure}
\emph{Phase-Locked (PL) lasers:} The PL-laser system basically
consists of two Extended Cavity Diode Lasers (ECDLs) called master
and slave laser (both characterized by narrow spectrum,
$\Gamma_{ECDLs}\approx0.5$ MHz). The master laser is stabilized to
the $^{87}$Rb D$_1$ line $5\,^2$S$_{1/2}$ F$_g=2 \rightarrow$
$5\,^2$P$_{1/2}$ F$_e=1$ transition, using an auxiliary evacuated
$^{87}$Rb cell in a DF-DAVLL configuration \cite{Wasik02}. Due to
the high speed servo loop for the laser stabilization, the linewdith
of the master is reduced by a factor $10$. The slave laser is - via
a heterodyne beat signal - phase locked to the master laser.
Therefore, the light beams from the master and the slave laser are
superimposed on a fast photodiode which detects the $6.8$ GHz beat
required for the experiments. After amplification, the heterodyne
beat signal is down converted in a double balanced ring mixer to 50
MHz and is compared with a reference frequency signal stemming from
an Intermediate Frequency (IF) oscillator. A phase/frequency
detector provides an output signal proportional to the phase
difference between the down converted beat-note and the
IF-Oscillator. To close the feedback loop the phase detector's
output signal is fed back to the slave laser. In the setup a
combined analogue-digital phase detector is used. In this way a
large capture range and a dead zone free locking is achieved
simultaneously. The root mean square (rms) phase noise level
(relative phase-jitter) of the PL-setup is $\Phi_{rms}\leq50$ mrad
(measured in the band of 1 Hz-1 MHz). As $\Phi_{rms}\ll\pi$ is well
fulfilled no degradation or additional broadening of the
CPT-resonance is observable \cite{Dalton82}. In the case of
PL-ECDLs, the frequency components of the dichromatic
electromagnetic field are separated by 6.835 GHz to bridge over the
splitting of the $^{87}$Rb hyperfine ground states. The intensities
as well as the linear polarization state are selected to be the same
for both frequency components. Finally, to avoid a residual
broadening of the CPT-resonances due to a wave vector mismatch, a
polarization maintaining single mode fibre is used to ensure perfect
collinear wave vectors of the two frequency components.

Figure \ref{fig:lockin-dc} shows the photocurrent variation (top plot) 
and the lock-in signal (bottom plot) for a
typical CPT resonance prepared with PL-ECDLs. In our experiments
the CPT resonance amplitude, $A_C$, is defined as the photocurrent variation relative to its background; 
and the CPT resonance linewidth, $\Delta\nu_C$, is the full width at half maximum (FWHM)
of the resonance. We chose this definition of $A_C$\cite{AC} because we aim the quantitative study
of quantum interference between dark states. Our approach allows
to separate the strength of the CPT effect, which is affected
by the quantum interference, and the
light level in the experiments, which changes using PL-ECDLs or
VCSEL because of the contribution of the off-resonance frequencies in
the modulated VCSEL spectrum. The role of the
(background) light level - mainly determining the Signal/Noise Ratio
- in the experiments is very important for the applications and
deserves to be studied in detail separately. In accordance with the definitions
given above, for the experimental conditions of figure
\ref{fig:lockin-dc} the normalized optical transmission due to the
CPT effect is about $4\%$ and $3\%$ for PL-ECDLs and modulated
VCSEL, respectively. 
%
\section{Influence of excited-state hfs on the lin$||$lin
CPT Resonance} \label{sec:result1}
%
%
The influence of the excited state hyperfine structure of the
$^{87}$Rb D$_1$ line is evidenced by studying the lin$\|$lin CPT
resonance versus the laser detuning ($\delta_L$) in a cell containing
$^{87}$Rb and N$_2$ as buffer gas, P$_{N_2}=0.5$ kPa. The
collisional broadening $\gamma_c$ at this pressure is about $ 2
\pi\cdot70$ MHz \cite{Romalis}, i.e. $\gamma_c$ is almost $12$ times
smaller than the splitting of the hyperfine excited states
($\omega_{HFS}=2 \pi\cdot817$ MHz). As a consequence the hyperfine
excited state structure remains resolved.

These measurements can be easily performed with a modulated VCSEL
since the VCSEL output frequency can be tuned with injection current
over a wide frequency range (more than ten GHz). The situation is
different for the PL-ECDLs. Here, the
frequency of the master laser (i.e. the detuning $\delta_L$) was
determined via an $^{87}$Rb saturation spectrum addi\-tio\-nal\-ly
overlapped by transmission fringes stemming from a confocal Fabry
Perot interferometer with a free spectral range of
$\nu_{FSR}=149.85$ MHz. Therefore a tuning range of about 2 GHz
(within the hfs of the excited state 5$^2$P$_{1/2}$) with a
frequency uncertainty of $\delta\nu\sim$ 5 MHz is achieved with both laser system.
\begin{figure}
  \begin{center}
    \resizebox{0.5\textwidth}{!}
    {
        \includegraphics{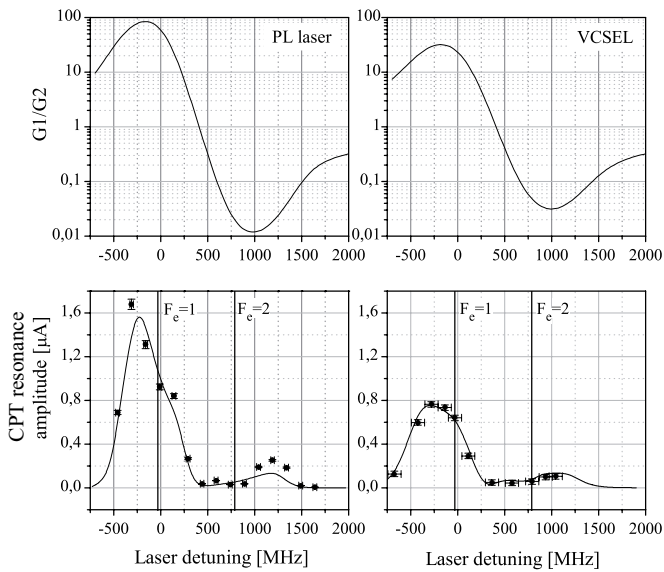}
    }
  \end{center}
  \caption{The lower plots show the dependence of CPT resonance amplitude (A$_C$)
  versus laser detuning ($\delta_L$) in case of a total resonant laser intensity
  of 3.8 mW/cm$^{2}$, a temperature of 68 $^{\circ}$C and for a Rb cell with P$_{N_2}=0.5$ kPa;
  while in the upper plots the G$_1$/G$_2$ ratio (calculated for both cases)
  is reported. The maximum of absorption is marked
  by the vertical lines (the pressure shift is about $-30$ MHz).
  The value $\delta_L=0$ is defined with respect to an evacuated $^{87}$Rb cell.
  In both cases, A$_C$ goes to zero when $(G_1/G_2)=1$ and has two maxima
  at the maximum and the minimum of $(G_1/G_2)$, respectively.
  Remark that the maxima of A$_C$ are shifted with respect to the maxima
  of absorption in the cell.}
  \label{fig:AVsD}
\end{figure}

The dependence of the CPT resonance amplitude (A$_C$) on $\delta_L$ is shown
in the lower plots of figure \ref{fig:AVsD}, for experiments
performed with PL-ECDLs and modulated VCSEL, left and right column
respectively. In figure \ref{fig:AVsD} the points are the results of
the experiments while the solid lines are the model results obtained
numerically by solving the coupled set of density matrix and Maxwell
equations with $\Gamma$ as free fit parameter (see section
\ref{sec:model}). The value of the ground state relaxation rate,
$\Gamma$, is determined by fitting the CPT resonance width $\Delta\nu_C$
obtained in the experiments. In case of the VCSEL source, a width
$\Delta\nu_C=1$ kHz and $\Delta\nu_C=10$ kHz is measured for a laser
detuning $\delta_L=0$ (transitions towards F$_e=1$) and
$\delta_L=+2\pi\cdot817$ MHz (transitions towards F$_e=2$)
respectively. As mentioned in \cite{Taichenachev05} the CPT
resonance arising from the group of transitions towards F$_e=2$ is
broader than the one arising from transitions towards F$_e=1$. But,
interestingly, negligible differences in the width $\Delta\nu_C$ are
observed for CPT resonances prepared by the PL ECDL system.

To quantify the influence of the each dark state contributing to the
CPT resonance, in section \ref{sec:model} the coefficients G$_e$ have
been introduced. In the two upper plots of figure \ref{fig:AVsD},
the calculated $(G_1/G_2)$ values are reported as a function of the
laser detuning for the two sets of data. In each plot $\delta_L=0$
refers to the un-perturbed group of transitions toward F$_e=1$
(evacuated cell); and the dotted lines represent the maxima of
absorption for F$_e=1$ ($\delta_L\approx-2\pi\cdot30$ MHz) and
F$_e=2$ ($\delta_L\approx +2\pi\cdot787$ MHz) in the cell containing
P$_{N_2}=0.5$ kPa (collisional shift is $(-22.2\pm0.4)$ GHz K
(kPa)$^{-1}$ \cite{Romalis}). For each laser system, we observe that
A$_C$ shows two relative maxima which are both shifted with respect
to the two maxima of absorption in the cell. In particular the
maxima of A$_C$ coincide with the ma\-xi\-mum and the minimum value
of $(G_1/G_2)$, respectively. Note that in general the dark state
$\left| \Psi'_a \right\rangle$ is affected by the losses due to the
one-photon transitions towards to the outermost excited levels
$5\,^2$P$_{1/2}$ F$_e=2$, m$_F=\pm2$ (see figure \ref{fig:linlin}).
For this reason when $(G_1/G_2)<1$ the resulting lin$\|$lin CPT
resonance is smaller than the one for $(G_1/G_2)>1$
\cite{Taichenachev05}. The values of A$_C$ noticeably decrease when
the light sources (PL-ECDLs or VCSEL) are tuned between the two
group of transitions. In fact we showed that the dark states
prepared via different excited hyperfine sublevels interfere
destructively, and, if their involvements into excitation process
are equal, i.e. $G_1/G_2=1$, the CPT resonance vanishes. Similar
effect has been discussed in the case of two Zeeman sublevels
belonging to the same hyperfine manifold in \cite{Happer67}.
Moreover, by comparing the left and right part of figure
\ref{fig:AVsD}, we can observe that the A$_C$ maximum recorded in
the experiments using PL-ECDLs is larger than the A$_C$ maximum
obtained with the modulated VCSEL. The measured behavior of A$_C$
vs. $\delta_L$ is well reproduced by the model calculations. In
particular - at the experimental conditions referring to figure
\ref{fig:AVsD} - the $G_e$ coefficients calculated for PL-ECDLs are
bigger than these coefficients calculated for the modulated VCSEL,
either for $e=1$ and $2$.

\begin{figure}
  \begin{center}
    \resizebox{0.45\textwidth}{!}
    { \includegraphics{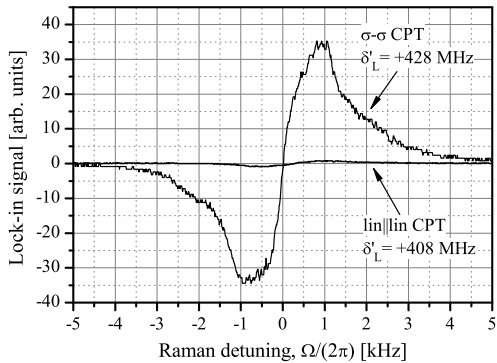}}
  \end{center}
  \caption{Evidence of the different nature of dark state excited with circular and linear light field,
  $\sigma$-$\sigma$ and lin$||$lin CPT resonance, respectively. The two resonances are recorded under the same experimental conditions.
  In particular, the laser (I = 3.8 mW/cm$^{2}$) is tuned near by the cross-over resonance
  ($\delta'_L=\delta_L/(2\pi)=+408$ MHz) of the $5\,^2$P$_{1/2}$ F$_e=1$ and F$_e=2$
  states (evacuated $^{87}$Rb cell). The lin$\|$lin CPT resonance is about $40$ times weaker than the $\sigma$-$\sigma$ one because of the destructive
  interference influence.}
  \label{fig:comp}
\end{figure}

The behavior of the $\sigma$-$\sigma$ CPT resonance amplitude, width
and position as a function of the laser detuning $\delta_L$ is
studied in a higher pressured buffer gas cell ($5.5$ kPa of Neon) by the
authors in \cite{Nagel}.
As anticipated in section \ref{intro} the dark states obtained by
interactions with linearly and circularly polarized are
intrinsically different. In the case of circularly polarized light
fields, the dark states arise from a vectorial coupling ($\Delta
m_F=0$) and no quantum interference occurs
\cite{Wynands98}. This different nature of the lin$||$lin and
$\sigma$-$\sigma$ CPT is shown in figure \ref{fig:comp}. Here,
the master laser (determining the laser detuning $\delta_L$) of the
PL-ECDLs is tuned at the crossover resonance between the
transitions: F$_g=1\leftrightarrow$F$_e$= 1 and
F$_g=1\leftrightarrow$F$_e$= 2. In case of the lin$||$lin CPT the
resonance amplitude $A_C$ is strongly suppressed because both dark states -
via the excited states F$_e$= 1 and F$_e$= 2 - have equal strengths
$G_1=G_2$ and interfere destructively (c.f. figure \ref{fig:AVsD}).
Unlike, in case of $\sigma$-$\sigma$ CPT such a destructive
interference is not observable due to the specific phase relation of
the Rabi frequencies.
\subsection{lin$||$lin CPT resonance amplitude versus homogeneous broadening}
\label{sec:result2}
\begin{figure}
  \begin{center}
    \resizebox{0.45\textwidth}{!}
    { \includegraphics{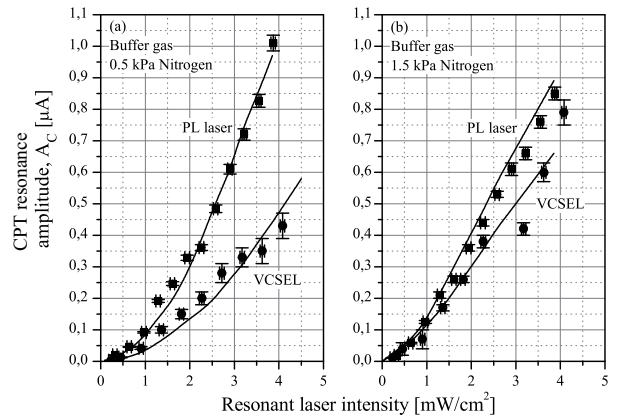}}
  \end{center}
  \caption{Dependence of lin$||$lin CPT resonance amplitude (A$_C$)
  versus the resonant laser power for experiments performed
  in two cells such as $P_{N_2}=0.5$ (a) and  $P_{N_2}=1.5$ kPa (b).
  The solid points refer to experiments and the lines show the numerical calculations.
  Remark that in (a) there is a noticeably difference in A$_C$ depending
  on the laser source, while in (b) that difference is reduced.}
  \label{fig:AVsI}
\end{figure}

The previous results allows us to predict the influence of the
homogeneous broadening $\gamma'$ (equation \ref{eq:gammaFirst}) on
the lin$\|$lin CPT resonance. In practice we compare the experiments
performed with PL-ECDLs and modulated VCSEL in a selected $^{87}$Rb cells
containing $0.5$ kPa of N$_2$ for a fixed value of the
laser detuning. In this section we compare the dark state prepared with the two lasers systems in
two $^{87}$Rb cell with different relevant pressure N$_2$. In particular we study the case of detuning
$\delta_L=0$, when the master laser of the PL-ECDLs is stabilized.
Under such conditions a PL-laser linewidth $\Gamma_{P}$ of
$2\pi\cdot0.04$ MHz is achieved, i.e. it is $2500$ times narrower
than the VCSEL linewidth ($\Gamma_V\approx2\pi\cdot100$ MHz).
Fi\-gu\-re \ref{fig:AVsI} shows the CPT resonance amplitude (A$_C$) versus the
resonant laser power for $^{87}$Rb cells with P$_{N_2}=0.5$ kPa and P$_{N_2}=1.5$ kPa,figure \ref{fig:AVsI}(a) and (b) respectively.
The main difference between the experiments performed with those cells is the relation of the
collisional broadening in each cell and the VCSEL linewidth. For
P$_{N_2}=0.5$ kPa, the collisional broadening contribution
$\gamma_{c}=2\pi\cdot70$ MHz is smaller than the linewidth of the
VCSEL ($\gamma_{c}<\Gamma_V$). On the contrary, for P$_{N_2}=1.5$
kPa $\gamma_c$ is $2\pi\cdot210$ MHz, i.e. about $2$ times of
$\Gamma_V$. In both cases, $\gamma_c$ is much bigger than the PL-laser linewidth
($\gamma_{c}\gg\Gamma_P$).
In figure \ref{fig:AVsI} (a) and (b) the solid points refer to the
experiments performed with the modulated VCSEL and the PL-ECDLs,
respectively. The solid lines show the results of the theoretical
model (c.f sec. \ref{sec:model}), which are in good quantitative
agreement with the experimental results. We observe that the
difference in A$_C$ obtained with the two laser systems is large in
(a) while it is reduced, and almost negligible, for I$_L<1$
mW/cm$^{2}$, in (b). To explain the observed behavior we first note
that in reference \cite{Affolderbach00} the authors showed a
reduction of the CPT resonance amplitude when the laser linewidth is larger
than the pressure broadening for the dark states created via the
interaction with circularly polarized laser light. They explained
this results by considering that the dark state is excited in atoms
belonging to several velocity classes. This argument is still valid,
of course, in the case of lin$\|$lin excitation, but can only
partially explain our results. In the previous parts of this
communication we have demonstrated that in the case of dark state
arising from a quadrupolar coupling ($\Delta m_F=2$) the excited
state hyperfine structure must be taken into account for each
$\delta_L$. Under quadrupolar coupling conditions the phase
relations of all Rabi frequencies, involved within the excitation
scheme, plays a major role. Using the formalism developed in our
model, we can conclude that for intensities lower than the
saturation of the optical transitions, when $\gamma_c
> \Gamma_L$ (condition satisfied in figure \ref{fig:AVsI} (b)), it
is possible to obtain the same $(G_1/G_2)$ ratio for the two laser
systems. It is possible, then, to find experimental conditions in
which the amplitude of the lin$\|$lin CPT resonance, A$_C$, is
independent of the laser linewidth. On the contrary, when
$\gamma_c<\Gamma_L$ (condition satisfied in figure
\ref{fig:AVsI}(a)), the laser linewidth plays an important role:
the ratio $(G_1/G_2)$ for experiments with the modulated VCSEL
is smaller than $(G_1/G_2)$ for the PL-ECDLs.
We can not appreciate the influence of the laser linewidth in the CPT
linewidth ($\Delta_C$). For both laser sources, $\Delta_C$ increases linearly with the laser
intensity in the same range of values. The values of $\Delta\nu_C$
extrapolated to zero laser intensities, are $\approx2\pi\cdot1.5$
kHz and $\approx2\pi\cdot0.8$ kHz for the cell with $0.5$ kPa and
$1.5$ kPa buffer gas pressure respectively, mainly caused by
collisional broadening effects \cite{Beverini}.
%
\section{Conclusions}
\label{sec:Concl} We present a study the CPT resonance prepared in
lin$\|$lin configuration, focusing our attention to the signal
between ground state Zeeman sublevels such as m$_F=\pm1$, which is a
good candidate for compact and high performance atomic clocks. This
interaction scheme is characterized by quantum interference between
dark states prepared through the two excited hyperfine states. A
model is developed taking into account the multi-level structure of
the atomic system and the linewidth of the lasers. We study the
signal as a function of the laser detuning, i.e. the excited state
hyperfine structure. By comparing the model and the experimental
results we can quantify the influence of each dark state on the
lin$\|$lin CPT resonance. Finally the good quantitative agreement
between theory and experiments allows predicting the effects of the
laser linewidth on the lin$\|$lin CPT resonance.
\section{Acknowledgments}
We are grateful to Prof. Pierre Thomann and Dr. Igor Mazets for
the sti\-mu\-la\-ting discussion. The work is supported by the
INTAS-CNES-NSAU grant 06-1000024-9321, by the the Fund for
Non-Profit 'Programs Dinastiya', by the Swiss National Research
Foundation (grant. 200020-118062), and by the Austrian Science
Found (FWF) (Project L300N02).


\begin{thebibliography}{}
\bibitem{AC} $A_C$ is proportional to the variation of the transmitted laser power through
the cell relative to a constant background.
\end{thebibliography}
\end{document}